\documentclass[final,1p,times]{elsarticle}

\usepackage{lineno,hyperref}
\usepackage{amsfonts, amsmath, booktabs, color}

\usepackage{algorithm}
\usepackage{algorithmic}

\journal{Computer Physics Communications}

\newcommand{\Fig}[1]{{\textbf{Figure~#1}}}

\newcommand{\Tab}[1]{{\textbf{Table~#1}}}

\newcommand{\p}{\partial}

\newcommand{\concentration}{{C}} 
\newcommand{\transmissibility}{{T}} 
\newcommand{\volume}{{V}} 
\newcommand{\saturation}{{S}} 
\newcommand{\pressure}{{P}} 
\newcommand{\porosity}{{\phi}} 
\newcommand{\perm}{{\kappa}} 
\newcommand{\elev}{{Z}} %
\newcommand{\adsorption}{{A}} 
\newcommand{\viscosity}{{\mu}} 
\newcommand{\ssvolrate}{{q}}
\newcommand{\reduction}{{R}} 
\newcommand{\rrf}{\mbox{RRF}} 
\newcommand{\wi}{\mbox{WI}} 

\newcommand{\speedup}{{s}}
\newcommand{\toddlong}{{\omega}} 

\newcommand{\phase}{{\alpha}} 
\newcommand{\aqueous}{{w}} 
\newcommand{\oleic}{{o}} 
\newcommand{\poly}{{p}} 
\newcommand{\widx}{{\varpi}} 
\newcommand{\lay}{{m}} 

\newcommand{\refer}{{0}} 
\newcommand{\xd}{{x}} 
\newcommand{\yd}{{y}} 
\newcommand{\zd}{{z}} 

\begin{document}

\begin{frontmatter}

\title{Numerical Simulations of Polymer Flooding Process in Porous Media on Distributed-memory Parallel Computers}

\author{He Zhong, Hui Liu, Tao Cui, Lihua Shen, Bo Yang, Ruijian He,
    Zhangxin Chen\corref{mycorrespondingauthor}}

\cortext[mycorrespondingauthor]{Corresponding authors: He Zhong, Hui Liu, Tao Cui, Zhangxin Chen} 
\ead{\{hzhong, hui.j.liu, lihua.shen, yang6, ruijian.he, zhachen\}@ucalgary.ca, tcui@lsec.cc.ac.cn}

\address[mymainaddress]{Chemical and Petroleum Engineering, Schulich School of Engineering, University of Calgary, Calgary T2N 1N4, Canada}

\begin{abstract}
Polymer flooding is a mature enhanced oil recovery technique that has been successfully applied in many field projects.
By injecting polymer into a reservoir, the viscosity of water is increased, and
    the efficiency of water flooding is improved. As a result, more oil can be recovered.
This paper presents numerical simulations of a polymer flooding process using parallel computers, where
    the numerical modeling of polymer retention, inaccessible pore volumes, a permeability reduction and polymer absorption
    are considered.
Darcy's law is employed to model the behavoir of a fluid in porous media, and
the upstream finite difference (volume) method is applied to discretize the mass conservation equations.
Numerical methods, including discretization schemes, linear solver methods, nonlinearization methods and parallel
techniques are introduced. Numerical experiments show that, on one hand, computed results match those from the commercial simulator,
    Schlumberger-Eclipse, which is widely applied by the petroleum industry, and, on the other hand,
    our simulator has excellent scalability, which is demonstrated by field applications with up to 27
    million grid blocks using up to 2048 CPU cores.
\end{abstract}

\begin{keyword}
High performance computing\sep Polymer flooding \sep Water flooding \sep Scalability 
\end{keyword}

\end{frontmatter}

\section{Introduction}

The emergence of parallel computers compels parallel computation techniques into an array of application areas,
including groundwater flow, contamination transport modeling, geothermal engineering, multiphase flow,
carbon dioxide sequestration and nuclear waste storage \cite{Coumou2008}. Beginning in the mid-1970s,
supercomputers were introduced to accelerate reservoir modeling problems through vectorization, and
computations could be completed at an advanced speed. However, models and programs had to be reorganized and
reworked to take advantage of leveraged computational power through vectorization. Besides, the program performance
deteriorated once a CPU number went beyond a specific number (usually 4, 8 or 16) \cite{Dogru2000}.  Except
on a shared memory system, parallel computations can also be carried out on distributed memory clusters.
Over the past few decades, significant progress has been made in developing high performance modeling tools for
distributed memory systems. However, they have not been widely applied in reservoir simulations. The reasons
are complicated but can be interpreted as follows: First, the need for high performance computing for a petroleum
reservoir simulation model is dominated by the physics of the underlying process. The severely nonlinear nature of a
physical process challenged the development of efficient parallel schemes for reservoir
simulations that result from the multiphase, multicomponent flow through heterogeneous porous media with
complex phase behavior equilibrium calculations. Second, the rapid advance in computing hardware strengthened
the traditional one-processor simulators, such as vectorization techniques in shared-memory machines
\cite{Coats1987}. It hindered the development of high performance parallel schemes, since PCs are much more
affordable and available than a cluster. The last but not the least, the earlier developed parallel algorithms
depended on a machine structure and were difficult to implement on different computers or architectures
\cite{Killiough1995,Wu2002}.

On the other hand, the demand for modeling capability has increased rapidly in recent years with an increase
in computational efforts. More complex geological, physical and chemical features are modeled through reservoir
simulations to assess new exploration and production technologies, such as enhanced recovery processes. In addition, the
traditional serial simulators have reached their simulation capability limits. The high performance simulation
technology has been progressively viewed as an important, alternative modeling approach to solve large-scale
simulation problems with multi-million and even multi-billion block models \cite{Edwards2012a}.

Simulation of multiphase flow involves the discretization of conservation and constraint equations
on a grid that is constructed from a set of blocks in the domain of interest. Finite difference methods (FDM) are
widely adapted by commercial software due to their simplicity and efficiency, while finite volume methods (FVM) are
also favoured by many engineering simulations due to their excellent conservation property and ability to handle complex geometry.
In either case, their
discretization procedure is designed for consistency, such that the errors incurred will vanish rapidly as
gridblock sizes approach zero. Nevertheless, the total number of grid blocks that can be
handled depends on the capacity of the available computing hardware and is necessarily limited. Clearly, this
constraint also applies to the need for resolution of small-scale phenomena. For example, the number of grid
blocks required to solve water front propagation phenomena in a secondary recovery model may not be
affordable in the context of a complete reservoir simulation and alternative strategies, such as an adaptive grid
system, may be required.

Initially, hydrocarbons are displaced from a reservoir by the natural reservoir energy, as the formation
pressure is considerably higher than the bottomhole pressure inside a wellbore. However, as the formation
pressure declines because of production, the recovery stage reaches its economic limit at a too low production
rate, or too high proportions of gas or water in the production stream. This stage is named the primary
production that produces only a small percentage of the initial hydrocarbons in place, typically around 10\%
for oil reservoirs. The second stage of hydrocarbon production maintains reservoir pressure and displaces
hydrocarbons toward a wellbore by injecting an external fluid, such as water or gas through injection wells.
However, when considerable amounts of injected fluid are produced from production wells, the production
is no longer economical. About additional 15\% to 40\% of the original oil can be recovered after the secondary recovery
method. Due to unfavorable mobility ratios and reservoir heterogeneities, early breakthrough during the
secondary recovery process prevents the injected water from sweeping the oil efficiently. However, it is
often not feasible to change the properties of the displaced fluid or the permeability to the
displaced fluid. Most mobility control processes of current interest involve addition of chemicals to the
injected fluid. These chemicals increase the apparent viscosity of the injected fluid and/or reduce the
effective permeability to the injected fluid. The chemicals used are primarily polymers when the
injected fluid is water \cite{Green1998}, which are the most popular choice due to their lower costs
compared to other types of additives. Other chemicals, such as surfactant or alkali, can also be injected
alternately or simultaneously to enhance oil recovery \cite{Lake1992}, but they are beyond the scope of this research.

By applying polymer, the viscosity of the water phase is increased, and, as a result, the mobility of the water
phase is reduced, which results in a more favorable fractional flow curve and then leads to a more efficient
sweeping pattern and reduced viscous fingering.  The mobility reduction of the injected water is due to two main
effects. First, the viscosity of a polymer solution is higher than that of pure water (the viscosity of
a polymer solution increases with raising polymer concentration). Second, the permeability to
water is reduced after the passage of a polymer solution through rock materials (the permeability to oil
is, however, largely unaffected).  Both effects reduce the water mobility while the oil mobility is unaltered.

Polymer flooding holds a bright future because it can improve the area swept efficiency not only in the macro
scale but also in the micro scale. The first polymer flooding application was reported in 1964
\cite{Pye1964,Sandiford1964}. The development of polymer flooding boomed in the US during the 1970s and 1980s
with several polymer flooding projects \cite{Lake1992}. However, it declined in the late 1980s because of low oil prices. During
the middle 1990s, polymer flooding was resumed in China to large extent. Especially, oil
production from polymer flooding contributed to 22.3\% of the total oil production in the Daqing oilfield by 2007
\cite{Wang2008,Saboorian-Jooybari2016}.  This significance has attracted the petroleum industry's interest in using reservoir simulators as
tools for reservoir evaluation and management to minimize operation costs and increase the process efficiency \cite{Chen2007b,Chen2006a}. Reservoir
simulators with special features are needed to represent coupled chemical and physical phenomena present in
polymer processes.

Bondor \cite{Bondor1972} presented the development of a three-phase, four-component, compressible, finite
difference polymer simulator. The model represented a polymer solution as a fourth component that was included
in the aqueous phase and was fully miscible with the water phase. Adsorption of polymer was represented as well
as the permeability reduction of the water phase. An implicit pressure-explicit saturation (IMPES) procedure was
used to solve the coupled system. Lutchmansingh \cite{Lutchmansingh1987} extended this simulator by solving
pressure and saturation distributions simultaneously and polymer concentration explicitly. Based on
Lutchmansingh's work, Abou-Kassem \cite{Abou-Kassem1996a} eliminated non-relevant equations and unknowns
by properly ordering the set of all equations and unknowns, thus providing significant savings in CPU time. Chang
\cite{Chang1990a} implemented a third-order finite difference method to capture a physical dispersion effect
which is normally smeared by artificial numerical dispersion. An IMPEC (implicit pressure-explicit concentration) scheme was adapted to solve an
isothermal, three-dimensional, miscible-flooding compositional model. The simulator is well known as {\it
UTCOMP}. On the other hand, its run time can be tremendous because of a timestep restriction on the IMPEC form, if a
large number of gridblocks are necessary for either simulation of a large reservoir or refinement of a model
for more accurate simulation. To overcome this computational limitation, a fully implicit formulation has been
adapted in the development of our simulator.

To capture fine-scale phenomena and optimize a polymer flooding process, large-scale reservoir
simulations with fine-scale grids are required. A parallel polymer flooding reservoir simulator
has been developed to address these issues.  In this paper, the mathematical model of polymer flooding is
introduced, including the conservation laws for water, oil and polymer, mechanisms of polymer flooding, and well
modelling. Numerical methods are presented. The upstream finite difference (volume) method is applied to
discretize the model equations. The standard Newton and inexact Newton methods are applied for their
highly nonlinear systems, and an algorithm for the inexact Newton method is introduced. Linear systems from
polymer flooding are ill-conditioned, especially when a reservoir has heterogeneous porosity
and permeability. In this case, the linear systems are difficult to solve. In our simulator, a multi-stage
preconditioner is employed to speed up the system solution. Parallel implementations are also introduced. Different
polymer flooding cases are used to illustrate the accuracy and scalability of our simulator.
The results show that this polymer flooding simulator has good scalability and large-scale reservoir models
can be simulated.


\section{Mathematical Model}

\label{sect:model}

The two-phase oil and water model is applied, and a temperature change is not considered here. The oil
component is assumed to stay in the oil phase, the water stays in the water phase, and polymer only distributes in the
water phase. The following sections will present a short summary of all related mathematical models for
rock, fluids and well handling.

\subsection{Rock Model}

When considering porous media at the macro-scale, the flow is governed by volume averaged equations. Each
computational block contains both solid and pore space which is filled with fluids, such as gas, oil and water.
The percentage of pore space, which is called porosity, is defined as
\[ \porosity = \frac{\volume_{pore}}{\volume_{bulk} } \]
where $\volume_{pore}$ is the volume of the pore space and $\volume_{bulk}$ is the volume of a block.
Porosity is a function of pressure (and temperature), and it can be modelled by the following equation:
\begin{equation}
    \phi(P) = \phi_r + c_r (P - P_r),
\end{equation}
where $c_r$ is the compressibility factor of the reservoir, $P$ is pressure, and $\phi_r$ is the reference porosity at the
reference pressure $P_r$.

\subsection{Fluid Model}
The notion of saturation $\saturation_\phase$ is introduced to define the
ratio of the volume of phase $\phase$ to the pore space in a block:
\[\saturation_\phase = \frac{\volume_\phase}{\volume_{pore}}. \]
The saturations of the oil phase ($\oleic$) and the water phase ($\aqueous$) satisfy the following relationship:
\begin{equation}
\saturation_\aqueous + \saturation_\oleic = 1.
\end{equation}
Darcy's law is applied to handle the relationship among flow rates of
a phase, reservoir properties, fluid properties and pressure in a reservoir, which is described as
\begin{equation}
Q = - \frac{\perm_e A\Delta \pressure}{\viscosity L},
\end{equation}
where $A$ is a cross-sectional area in a flow direction,  $\Delta \pressure$ is a pressure difference, $\viscosity$
is the viscosity of a fluid, and $L$ is the length of a porous medium in the flow direction.
$\perm_e$ is the effective permeability for the given phase, which is the product of absolute permeability $\perm$ and relative permeability $\perm_r$. $\perm$ is defined as a tensor with respect to all the $\xd$, $\yd$ and
$\zd$ directions; mostly, it is a diagonal tensor: $\perm = (\perm_\xd, \perm_\yd, \perm_\zd)$.  Darcy's law can also be rewritten, with Darcy's
velocity $\ssvolrate$,
\begin{equation}
\ssvolrate = \frac{Q}{A} = - \frac{\perm_e}{\viscosity} \nabla \pressure.
\end{equation}
With gravity, the mass of each phase satisfies the following conservation law \cite{Chen2007}:
\begin{equation}
\frac{\p}{\p t} \left(\porosity \saturation_\phase \rho_\phase \right)  = \nabla\cdot\left(\frac{\perm
    \perm_{r\phase} \rho_\phase}{\viscosity_\phase}\left(\nabla\pressure_\phase - \gamma_\phase \nabla\elev\right)\right) + \ssvolrate_\phase, \qquad \phase = \aqueous, \oleic
\end{equation}
where $\rho_\phase$ is the phase density, 
$\ssvolrate_\phase$ is the source term that models the mass changes caused by injection or production wells,
 $\gamma$ is the gravity, $\elev$ is the depth of a block, and
$\perm_{r\phase}$ stands for the relative permeability for the $\phase$ phase.
In addition, when polymer exists in the water phase, the mass conservation law for the water phase becomes
\begin{equation}
\frac{\p}{\p t} \left(\porosity \saturation_\aqueous \rho_\aqueous \right)  = \nabla\cdot\left(\frac{\perm
    \perm_{r\aqueous} \rho_\aqueous}{R_k\viscosity_{\aqueous,e}}\left(\nabla\pressure_\aqueous - \gamma_\aqueous \nabla\elev\right)\right)
    + \ssvolrate_\aqueous
\end{equation}
where $\reduction_k$ is the permeability reduction factor caused by polymer and $\viscosity_{\aqueous,e}$ is the
viscosity of a water-polymer solution. The definitions of $\reduction_k$ and $\viscosity_{\aqueous,e}$ will be
introduced later.

The water phase pressure, $\pressure_\aqueous$, and the oil phase pressure, $\pressure_\oleic$, are related by
\begin{equation}
    \pressure_c(\saturation_\aqueous) = \pressure_\oleic - \pressure_\aqueous.
\end{equation}
The pressure difference is called the capillary pressure, which usually depends on the saturations of the phases in porous media and
is measured by lab experiments. If saturation and any phase pressure are known, the other phase pressure can be calculated by the above formula.

\subsection{Polymer Model}

The flow of polymer is assumed to act as a component dissolved in the water phase, which is modeled by the following equation:
\begin{equation}
    \frac{\p}{\p t}\left({\porosity \saturation_\aqueous \rho_\aqueous \concentration_\poly} +
    (1-\porosity) \adsorption_d \right)  = \nabla\cdot\left(\frac{\rho_\aqueous \concentration_\poly \perm
    \perm_{r\aqueous}}{R_k \viscosity_{p,e}}\left(\nabla\pressure_\aqueous - \gamma_\aqueous \nabla\elev\right) \right)
    + \ssvolrate_\aqueous \concentration_\poly
\end{equation}
where $\concentration_\poly$ is the concentration of the polymer in the water phase and $\adsorption_d$ is the
polymer adsorbed by the reservoir.

When polymer molecules flow through porous media, part of them are restricted in pores, where only
water or brine is allowed to pass by with a reduced mobility.
As the polymer solution interacts with the reservoir rock, polymer is adsorbed or desorbed from the rock
surface; this mechanism is known as \emph{polymer retention}. There are two mechanisms during the polymer
retention process, which are separated as adsorption of the polymer on rock surfaces and entrapment of polymer
molecules in small pore space. Both these mechanisms increase the resistance of flow. These effects are
modeled by reducing the permeability of the rock to water.

The long chains of polymer molecules can flow into a large pore opening and get trapped when
the other end has a smaller opening. Entrapment can also take place when the flow is restricted or stopped.
When this happens, the polymer molecules lose their elongated shape and coil up. Desorption of the polymer from
the reservoir rock can also take place if sufficient polymer has already been adsorbed above a residual
sorption level. It is difficult to quantify what percentage of injected polymer is adsorbed and what
percentage is trapped in small pore spaces since only the produced polymer concentration can be measured. Both
these mechanisms result in a loss of polymer to the reservoir.

The adsorption process causes a reduction in the permeability of the rock to the passage of the aqueous phase
and is directly correlated to the adsorbed polymer concentration.
The reduction factor, $\reduction_k$, is a
function of polymer adsorption and the \emph{residual resistance factor} ($\rrf$), which is expressed as
\begin{equation}
\reduction_k = 1.0 + (\rrf - 1.0) \frac{\adsorption_d}{\adsorption_{d,max}}
\end{equation}
where $\adsorption_d$ is the cumulative adsorption of polymer per unit volume of the reservoir rock and
$\adsorption_{d,max}$ represents the maximum value of $\adsorption_d$, which denotes the maximum adsorptive
capacity of polymer per unit volume of the reservoir rock. Both $\rrf$ and $\adsorption_{d,max}$ are functions
of the reservoir rock permeability.

Assuming equilibrium sorption with the reservoir rock, the sorption phenomenon can be described as a function
of polymer concentration $\concentration_\poly$ only:
\begin{equation}
\adsorption_d = f(\concentration_\poly)
\end{equation}
This relationship is specified in the form of a table.

Not only is the rock permeability to water reduced after the passage of a polymer solution through porous
media, but also the viscosity of the polymer solution is higher than that of pure water. Small
concentrations of polymer, on the order of a few hundred to a few thousand ppm (by weight), increase the
viscosity of an aqueous solution significantly \cite{Green1998}.

The \emph{Todd-Longstaff technique} is used to calculate the effective viscosity that incorporates the effect of
physical dispersion at the leading edge of a slug and also the fingering effect at the rear edge of the
slug \cite{Lie2012}. The viscosity of a fully mixed polymer solution, denoted by
$\viscosity_m(\concentration_\poly)$, rises as the polymer concentration ($\concentration_\poly$) increases.
The viscosity of the solution at the maximum polymer concentration is also specified that is denoted by
$\viscosity_\poly^\refer$. Then the effective polymer viscosity is taken to be \begin{equation}
\viscosity_{\poly, e} = \left(\viscosity_m(\concentration_\poly)\right)^\toddlong
\left(\viscosity_\poly^\refer\right)^{1-\toddlong} \end{equation} where $\toddlong$ is the Todd-Longstaff
mixing parameter. The mixing parameter is useful in modeling the degree of segregation between water and the
injected polymer solution. If $\toddlong = 1$,  then the polymer solution and water are fully mixied. If
$\toddlong = 0$, the polymer solution is completely segregated from the water.

A partially mixed water viscosity is calculated in an analogous manner using the fully mixed polymer
viscosity and the pure water viscosity \begin{equation} \viscosity_{\aqueous, partial} =
\left(\viscosity_m(\concentration_\poly)\right)^\toddlong\left(\viscosity_\aqueous\right)^{1-\toddlong}
\end{equation} The effective water viscosity is calcualted by the partially mixed water viscosity and the effective
polymer viscosity as a harmonic average: \begin{equation} \frac{1}{\viscosity_{\aqueous, e} } =
\frac{\alpha}{\viscosity_{\poly,e} } + \frac{1-\alpha}{\viscosity_{\aqueous, partial} } \end{equation}
where $\alpha$ is the effective saturation of the injected polymer solution within the total aqueous phase in a
block.

The mixing of polymer and water modifies the solution viscosity as well. Since polymer has higher viscosity
compared to pure water, no matter which mixing rule is selected, the mixture viscosity increases as a function of
polymer concentration in the solution. Two commonly used mixing rules are used, which include a linear mixing
rule: \begin{equation} \bar{\viscosity}_\aqueous = \beta \viscosity_\poly^\refer +
(1-\beta)\viscosity_\aqueous \end{equation}
and a nonlinear mixing rule: \begin{equation} \bar{\viscosity}_\aqueous = \left( \viscosity_\poly^\refer
\right)^{\beta}  \left( \viscosity_\aqueous \right)^{1-\beta} \end{equation}
where $\beta$ is a parameter dependent on polymer concentration given by \begin{equation} \beta =
\frac{\concentration_\poly}{\concentration_\poly^\refer} \end{equation}

A higher water viscosity and a reduction in permeability will result in an increase in the resistance to flow,
and divert the polymer solution toward
areas unswept by water. This mechanism is well-known as \emph{mobility control}. Directly, the water-oil
mobility ratio is reduced to close to unity or less. Then the volumetric sweep efficiency is improved and
higher oil recovery is achieved compared to conventional water flooding.

As mentioned above, polymer molecules can flow into large pore openings. However,
there are also small openings which are not contacted by polymer molecules.
To describe this phenomenon, an \emph{inaccessible pore volume} (IPV) is used to measure all the pore space that may not be
accessible to polymer molecules. The presence of IPV causes the polymer solution to travel at a greater velocity than inactive tracers embedded
in water. This chromatographic effect is modeled by assuming that the IPV is constant for each rock type
and either does not exceed the corresponding irreducible water saturation or is independent of the water
saturation.  The concept of IPV allows a polymer solution to advance and displace oil at a faster
rate than predicted on the basis of total porosity.

\subsection{Wellbore Models}

A numerical simulation of fluid flows in petroleum reservoirs must account for the presence of wells.  They
supply a set of realistic boundary conditions for computations of pressure distributions \cite{Dumkwu2012}.
The fundamental task in modeling wells is to model flows into/from a wellbore accurately and to develop
accurate well equations that allow the computation of the bottom hole pressure with a given production or
injection rate, or the computation of a rate with known pressure \cite{Chen2006a}.

Peaceman \cite{Peaceman1978,Peaceman1983a} associated a steady-state pressure for an actual well with the
computed pressure at a grid block through the concept of an equivalent radius $r_e$.
If a well was completed in more than one grid block, a well index ($\wi$) was introduced to account for
well pressure losses within the grid blocks due to the radial inflow into the well. The well index depends on
the geometry of a grid block, location and orientation of the well segment in that grid block, anisotropic
reservoir property and a skin factor \cite{STARS}:
\begin{equation}
    \label{eq:wi}
    \wi = \frac{2\pi f h f_h \perm_a}{\ln(r_e/r_w) + s}
\end{equation}
where $f$ is the well fraction that is evaluated by the angle open
to flow and varies due to the well position in a grid block. It equals 1 for a well going approximately
through the center of a grid block. $h$ represents a grid block thickness along the well direction, and
$f_h$ is the grid block thickness factor.  The current completion length in the current grid block is the
product of $h$ and $f_h$. $\perm_a$ is the geometric average permeability and estimates the
formation's absolute permeability perpendicular to the well direction. $s$ denotes the skin factor, which may
also differ from one perforated block to another within the wellbore. This is especially true if different
perforation densities and intervals exist within each individual simulation layer.  $r_w$ is the wellbore
radius. The equivalent radius $r_e$  and formation absolute permeability $\perm_a$ are computed according to the
wellbore direction and the discretization procedures. For instance, if a well is parallel to the $\xd$-direction
in a Cartesian grid, then
\begin{equation}
    \label{eq:wi:re:k}
    \begin{aligned}
        r_{e} &= \frac{2 g_f}{\sqrt{\pi}}
        \frac{(\sqrt{\perm_{\zd}/\perm_{\yd}}h_\yd^2 +\sqrt{\perm_{\yd}/\perm_{\zd}}h_\zd^2)^{1/2}}
        {(\perm_{\zd}/\perm_{\yd})^{1/4} + (\perm_{\yd}/\perm_{\zd})^{1/4} } \\
        \perm_a &= \sqrt{\perm_{\yd}\perm_{\zd}}
    \end{aligned}
\end{equation}
Similar to the well fraction $f$, the factor $g_f$ in ({\ref{eq:wi:re:k}) depends on the geometry of a
grid. It equals 0.249 for a well going approximately through the center of a grid  block. Detailed
information can be found in \cite{STARS,Palagi1994a}.

The flow rate, $q_{\lay, \phase}$, for the $\phase$-phase in a perforated grid block $\lay$ is the product of
the well index, the fluid mobility and the drawdown pressure \cite{Chen2007, Fung2005}:
\begin{equation}
    \label{eq:ssw:Q}
    q_{\lay,\phase} = \wi_\lay  \lambda_\phase \rho_\phase \left(\pressure_{b,_\lay} - \pressure_\lay \right)
\end{equation}
The wellbore pressure at each grid completion ($\pressure_{b,_\lay}$) is
different from one layer to another, depending on the existing pressure drop in a wellbore. It is
calculated by the hydrostatic pressure difference drawn from the average density of the fluid mixture
in the wellbore:
\begin{equation}
    \label{eq:ssw:dp}
    \pressure_{b,_\lay} = \pressure_{b} + \gamma_{well} (z_\lay - z_{b})
\end{equation}
where $\gamma_{well}$ is the fluid unit weight which depends on the fluid mixture
density in the wellbore and $\pressure_{b}$ is the reference bottom hole pressure at reference depth $z_b$
\cite{Coats1980, Fung2005}.

To optimize oil production and to reduce operation costs, various well operations may be employed at any time,
such as fixed bottom hole pressure, a fixed oil production rate, a fixed water production rate, a
fixed water injection rate, or a fixed liquid production rate.
When the fixed bottom hole pressure well operation is applied to a well, the constraint for the well is
described as
\begin{equation}
\pressure_b = c,
\end{equation}
where $c$ is a constant.
The fixed water rate condition is the following equation:
\begin{equation}
    \sum_{m} {q_{m,w}} = q_{c,w},
\end{equation}
where $q_{c,w}$ is a constant. For the fixed oil rate production operation, the constraint is
\begin{equation}
    \sum_{m} {q_{m,o}} = q_{c,o},
\end{equation}
where $q_{c,o}$ is a fixed constant. For the fixed liquid production rate operation, the production of oil and
water is fixed, and its constraint equation is
\begin{equation}
\sum_{m} ({q_{m,o} + q_{m,w}}) = q_{c,l},
\end{equation}
where $q_c$ is a constant.

\section{Numerical Methods and Parallelization}

The reservoir model for polymer flooding is highly nonlinear, which is hard to solve analytically.
In this paper, numerical solutions are obtained by the fully implicit method thanks to its unconditional
stability. \Fig{\ref{fig:pflow:gdd}} shows the details of the solution flowchart, which includes the following
steps:
\begin{enumerate}
    \item Loading model. In this step, a model file is loaded, which contains
        information for reservoirs, such as permeability, porosity and geometry,
        for oil, water and polymer, such as density, viscosity, and concentration,
        for well operations, and for numerical parameters. The model may have hundreds of parameters.

    \item Grid generation and distribution. The model file defines a grid, such as dimensions in the $x$, $y$,
        and $z$ directions, sizes of each gridblock, and coordinates. Since the simulation is parallel,
        the grid must be distributed to each MPI, and a communication structure must be set up.

    \item Initialization. This step sets the initial pressure, saturations of oil and water, concentration of
        polymer, bottom hole pressure, porosity and permeability of the reservoir, and other properties, such
        as relative permeability, density, viscosity, and well index.

    \item Time discretization and time step selection. A time step is dynamically selected, which satisfies
        some conditions:
        \begin{enumerate}
            \item If Newton methods fail, the time step will be cut.

            \item In one time step, well operations keep unchanged.

            \item The time step has a maximal value, which is set by the model file.

            \item The algorithm always attempts to increase a time step to reduce simulation time.

        \end{enumerate}

    \item Newton iteration. In each time step, an nonlinear system is solved by the Newton method, which converts
        an nonlinear system to a linear system, $J x = b$. Inside this step, the properties for the reservoir and
        fluids must be computed, such as porosity, density and viscosity.

    \item Linear iteration. This step solves the linear system $J x = b$. A proper linear solver, a preconditioner
        and solution parameters must be chosen, which can be input by the model file.

\end{enumerate}

\begin{figure*}[!hbt]
\centering
\includegraphics[width=0.8\textwidth]{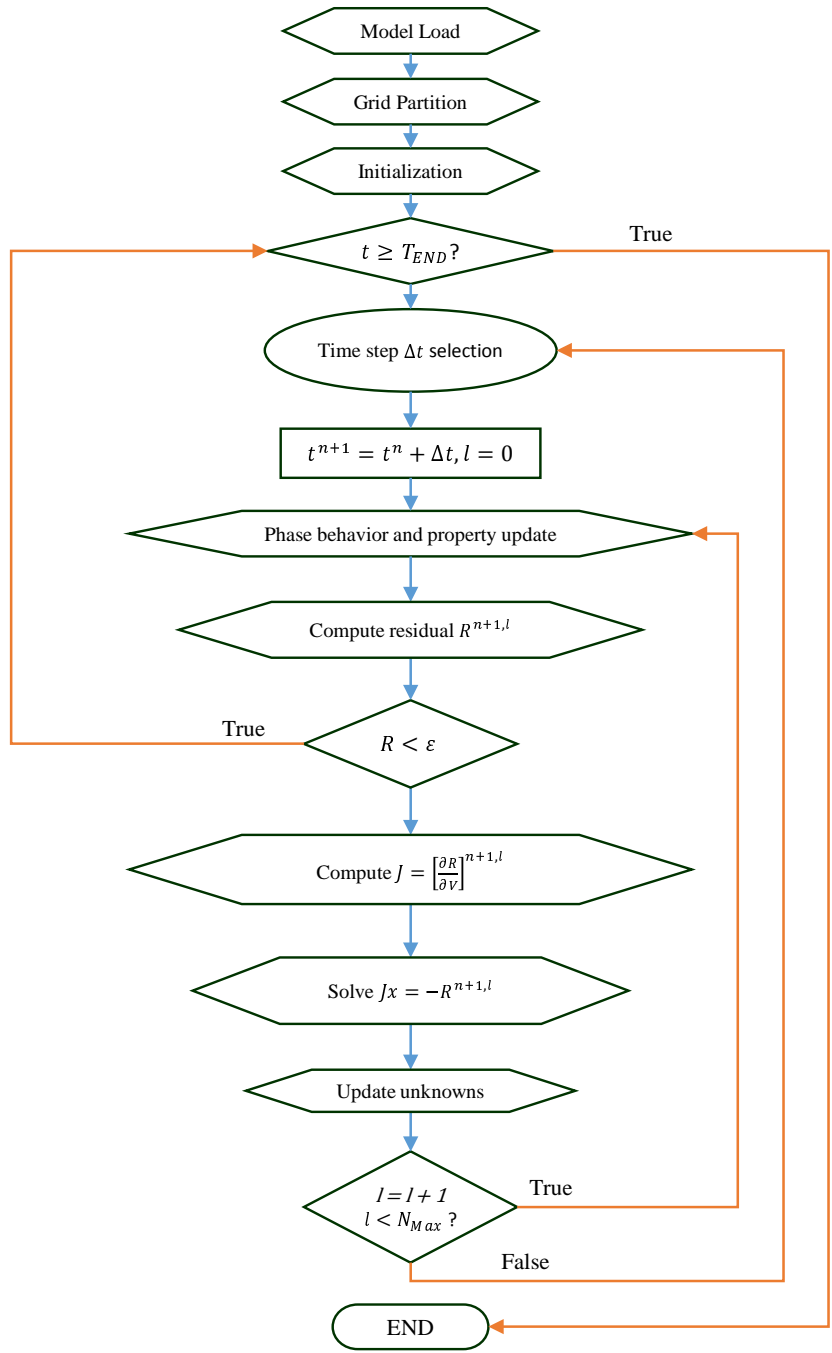}
\caption{Parallelism protocol of general domain decomposition strategy.}\label{fig:pflow:gdd}
\end{figure*}

The MPI (Message Passing Interface) is applied to handle communication among computation nodes. When
calculating properties, such as transmissibility, neighboring information is always required. To develop a
scalable parallel application, communications should be minimized. In reservoir simulations, the communication
pattern is determined by a grid distribution.

Grid partitioning algorithms aim to minimize the idle time and communication flow between different processors
by dividing the blocks equally among partitions and minimizing the number of partition spanning edges.  The
quality of the partitioning plays a crucial role in the performance of applications.
Reservoir simulation applies grid-based numerical methods, such as the finite volume and finite difference
methods, to approximate a governing system. Grid blocks have a higher possibility to communicate with
each other when they are closer to each other. Based on this assumption, we have introduced a modified Hilbert
space-filling curve partitioning method \cite{Liu2017a} in our in-house simulator, which has shown promising
loading balance and excellent scalability.

After the grid partitioning subroutine occurs, the subregion topology graph and block level connection lists
are set up. The block level connection lists record the geometry information
both in the global and local indices. Here, the global index is normally defined by the natural ordering. It
labels the grid blocks in the $\xd$, $\yd$ and $\zd$ directions, respectively. Once the subregion configuration
is fixed, the dataset is read and scattered into each processor. A parallel IO algorithm through MPI-IO is
implemented to read the data files.

A set of data structures have been designed to store distributed data, such as DOF (Degrees of Freedom), VEC,
MAT, SOLVER, and SOLVER\_PC \cite{prsi}. A DOF is defined on a grid, which can be defined to store
block-based data, such as porosity, density and viscosity, and on a well, which can be defined for bottom hole
pressure, a well rate, and a well index. The VEC and MAT are distributed vectors and matrices. The reader refers
to our previous paper \cite{prsi} for more details of parallelization.

\subsection{Time Discretization}

In reservoir simulation, to study well performance and optimize production, a simulation period is usually long,
such as over 10 years. Here the backward Euler method is applied to discretize a time derivative. Let $u$
be a function,
$u^n$ be the solution of $u$ at a time step $n$, and $F$ be an nonlinear system. For the following
differential equation:
\begin{equation}
    u_t = F(u, t),
\end{equation}
the backward Euler method computes the numerical solution by
\begin{equation}
    (\frac{\partial u}{\partial t})^{n+1} = \frac{u^{n+1} - u^{n}}{\Delta t} = F(u^{n + 1}, t^{n+1}),
\end{equation}
where $\Delta t$ is a time step. This method is implicit, and the above system is still nonlinear,
which must be solved at each time step.

\subsection{Spatial Discretization}

The oil phase pressure, water saturation, polymer concentration and well bottom hole pressure are chosen as
the primary unknowns, and other unknowns are functions of these primary unknowns.

When fluids move in a reservoir, there may be fluid exchange in two neighboring blocks. The term,
transmissibility, is defined to describe the amount of fluid exchange.
Here, let $d  ~ (d = \xd, \yd, \zd)$ be any space direction and $A$ be the area of a face in the $d$ direction;
then the transmissibility term $\transmissibility_{\phase,d}$ for phase $\phase$ is defined as
\begin{equation}
\transmissibility_{\phase,d} = \frac{\perm \perm_{r\phase}}{\reduction \viscosity_{\phase}} \rho_{\phase} \frac{A}{\Delta d},
\end{equation}
where $\Delta d$ is the cell length along the $d$ direction, $\perm$ is the permeability, $\perm_{r\phase}$ is the relative
permeability, $\viscosity_\phase$ is the viscosity, and $\reduction$ is 1 for the oil phase and $\reduction_k$ for water and polymer.

The transmissibility is defined on each face of a block. For any two neighboring blocks, since they share a
face, the value of the transmissibility term is the same for these two blocks.
Different weighting schemes must be applied to average different properties at an interface for them to make a physical sense.
The geometric properties, such as a cell length $\Delta d$ and a cross area $A$, and rock permeability $\perm$ are harmonically averaged. Since the fluid properties, such as $\viscosity_\phase$ and $\rho_\phase$, do not change much, they are averaged by a pore volume-weighted arithmetic average scheme. An upstream weighting technique is applied to calculate a relative permeability, which means that a relative permeability at the interface of two neighboring blocks is calculated from the block that has a higher potential.
For example, the relative permeability $\perm_{r\phase}$ at the interface $(i-1/2, j, k)$ is defined as
\begin{equation}
(\perm_{r\phase})_{i - \frac{1}{2},j,k} =
\left\{
\begin{aligned}
& (\perm_{r\phase})_{i,j,k} & \mbox{ if  }  \Phi_{i,j,k} \ge \Phi_{i-1,j,k} \\
& (\perm_{r\phase})_{i-1,j,k} & \mbox{ if  }  \Phi_{i,j,k} < \Phi_{i-1,j,k}
\end{aligned}
\right..
\end{equation}
Other higher-order upstream weighting techniques can also be used for relative permeabilities \cite{Chen2007}.
\subsection{Newton Methods}

The standard Newton method is the usual way to solve nonlinear systems. In our implementation, an inexact Newton
method \cite{inexact-Newton} is also applied, which relaxes the termination tolerance to reduce computation time.
Its algorithm is described in Algorithm \ref{inewton-alg}. The only difference from the standard Newton method is
that the latter uses a fixed termination tolerance for a linear solver while the inexact Newton method
uses a dynamic tolerance, which is computed automatically by equation (\ref{fterm}).

\begin{algorithm}[!htb]
\caption{The Inexact Newton Method}
\label{inewton-alg}
\begin{algorithmic}[1]
    \STATE Choose an initial solution $x^0$ and a termination tolerance $\epsilon$.
    \STATE Assemble the right-hand side $b$ and set $l = 0$.
    \WHILE{$\left\|b \right\| \ge \epsilon$}
    \STATE Assemble the Jacobian matrix $J$.
    \STATE Determine linear solver termination tolerance $\eta_l$.
    \STATE Find solution $x$ such that
    \begin{equation}
        \left\| J x - b \right\| \leq \eta_l \left\| b \right\|,
    \end{equation}
    \STATE    $x^l = x^{l-1} + x$.
    \STATE    Let $l = l+1$.
    \ENDWHILE
    \STATE $x^* = x^l$ is the solution of the nonlinear system.
\end{algorithmic}
\end{algorithm}

$\eta_l$ is the dynamic termination tolerance for the linear system $J x = b$.
The choice of this parameter is designed to speed the convergence of the Newton method, to avoid
over-solution of a linear system, and to reduce computation time. Three popular algorithms are provided as
follows:
\begin{equation}
    \label{fterm}
    \eta_l =
    \left\{
        \begin{aligned}
            & \frac{\left\| b^l - r^{l-1} \right\|}{\left\| b^{l-1} \right\|}, \\
            & \frac{\left\| b^l \right\| - \left\| r^{l-1} \right\|}{\left\| b^{l-1} \right\|}, \\
            & \gamma \left( \frac{\left\| b^l \right\|}{\left\| b^{l-1} \right\|} \right)^{\beta},
        \end{aligned}
        \right.
\end{equation}
where $r^l$ is the residual vector of the $l$-th Newton iteration and $b^l$ is the right-hand side vector.
The third formula is applied in our simulator, where $\gamma$ is 0.5 and $\beta$ is $\frac{\sqrt{5} + 1}{2}$.
$\eta_l$ is also tailored to meet this contition: $\eta_l \in [0.01, 0.1]$.

\subsection{Linear Solver}

A Jacobian matrix is nonsymmetric and highly ill-conditioned. The Krylov subspace solvers are applied to
solve the linear system $J x = b$. In real application, a preconditioner $M$ is always applied to solve an
equivalent linear system $M^{-1} J x = M^{-1} b$. A family of scalable CPR methods \cite{CPR-H} have been developed to handle
linear systems from reservoir simulations, and the CPR-FP method in \cite{CPR-H} is used as the preconditioner.

The system has four unkonws, oil phase pressure ($\pressure_\oleic$), water saturation ($\saturation_\aqueous$), polymer concentration
($\concentration_\poly$) and well bottom hole pressure $(\pressure_b$), which are vectors. There are two common strategies to arrange
the unknown $x$, which are point-wise and block-wise, respectively. For the point-wise strategy, $x$ is written as
\begin{equation}
    {x} =
    \left(
    \begin{array}{c}
        {\pressure_\oleic} \\
        {\saturation_\aqueous} \\
        {\concentration_\poly} \\
        {\pressure_b}
    \end{array}
    \right),
\end{equation}
while, for the block-wise strategy, $x$ is written as
\begin{equation}
    x = \left(
    \begin{array}{c}
        \pressure_{\oleic,1} \\
        \saturation_{\aqueous,1} \\
        \concentration_{\poly,1} \\
        \cdots  \\
        \pressure_{\oleic,n} \\
        \saturation_{\aqueous,n} \\
        \concentration_{\poly,n} \\
        \pressure_{b,1} \\
        \cdots  \\
        \pressure_{b,n_\widx} \\
    \end{array}
    \right),
\end{equation}
where $n$ is the number of grid blocks and $n_\widx$ is the number of wells. In real simulations, the preconditioner
using the block-wise strategy has better convergence than that using the point-wise strategy. However, if the
point-wise strategy is applied, the Jacobian matrix J has a clear block structure as
\begin{equation}
\label{mat-ab}
J = \left(
        \begin{array}{llll}
            J_{\pressure\pressure} \hspace{0.1cm}   & J_{\pressure\saturation} \hspace{0.1cm} & J_{\pressure\concentration} \hspace{0.1cm} & J_{\pressure\widx}   \\
            J_{\saturation\pressure} \hspace{0.1cm}   & J_{\saturation\saturation} \hspace{0.1cm} & J_{\saturation \concentration} \hspace{0.1cm} & J_{\saturation \widx}   \\
            J_{\concentration\pressure} \hspace{0.1cm}   & J_{\concentration\saturation} \hspace{0.1cm} & J_{\concentration\concentration} \hspace{0.1cm} & J_{\concentration\widx}   \\
            J_{\widx\pressure} \hspace{0.1cm}   & J_{\widx\saturation} \hspace{0.1cm} & J_{\widx \concentration} \hspace{0.1cm} & J_{\widx\widx}   \\
        \end{array}
        \right),
\end{equation}
where $J_{\pressure\pressure} \in R^{n \times n}$ is the matrix corresponding to the oil phase pressure,
$J_{\saturation\saturation} \in R^{n \times n}$ is the matrix corresponding to the water saturation, $J_{\concentration\concentration} \in R^{n \times n}$
is the matrix corresponding to the polymer concentration, $J_{\widx\widx} \in R^{n_\widx \times n_\widx}$ is the matrix
corresponding to the well bottom hole pressure, and other matrices are coupled terms.

If we define a restriction operator from $x$ to $\pressure_\oleic$, then its formal formula can be written as
\begin{equation}
\varPi_r  x = \pressure_\oleic.
\end{equation}
A prolongation operator $\varPi_p$ from $\pressure_\oleic$ to $x$ can be defined as
\begin{equation}
    \varPi_p  \pressure_\oleic = \left(
    \begin{array}{c}
        \pressure_{\oleic,1} \\
        \cdots   \\
        \pressure_{\oleic,n}  \vspace{0.1cm} \\
        \overrightarrow{0} \vspace{0.1cm}  \\
        \cdots \vspace{0.1cm} \\
        \overrightarrow{0} \vspace{0.1cm} \\
        \overrightarrow{0} \\
    \end{array}
    \right).
\end{equation}

The preconditioning linear system $M y = f$ must be solved in each iteration.
The CPR-FP method \cite{CPR-H} can be described by Algorithm \ref{pc-fp}, where the first step is to solve an
approximate solution using restricted additive Schwarz (RAS) method and the third step is to solve the subproblem by
algebraic multigrid method (AMG). It is well known that RAS method and AMG method are scalable for parallel
computing, so the CPR-FP method is also scalable.

\begin{algorithm}[!htb]
    \caption{The CPR-FP Method}
    \label{pc-fp}
    \begin{algorithmic}[1]
        \STATE $y = D_r(J)^{-1} f$
        \STATE {$r = f - Jy$}
        \STATE $y = y + \varPi_p M_g(J_{\pressure\pressure})^{-1} \varPi_r  r $
    \end{algorithmic}
\end{algorithm}

\section{Numerical Experiments}

\subsection{Model Validation}

In this section, we present several results to validate the application of our simulator (BOS) in different polymer
flooding projects. First, a homogeneous, one-dimensional polymer flooding case with 15 blocks is presented. The polymer
concentration varies in different production stages. Second, we consider a two-dimensional polymer flood
with multiple constraints on the injection wells. Finally, we consider a three-dimensional case with a
stratified reservoir. It has a number of layers with significant or complete lateral continuity. The layer
permeabilities vary greatly and adjacent layers communicate vertically.
Simulation results show that water advances rapidly in highly permeable layers and slowly in tight layers as the driving fluid to displace oil.

ECLIPSE100, Schlumberger, is a fully-implicit, three-phase, three-dimensional, general purpose black oil
simulator that can model different chemical EOR (enhanced oil recovery) processes, including polymer and surfactant flooding.
It can be run in fully implicit and adaptive implicit modes, and is widely considered as a reference and benchmark standard, which has been successful in reproducing laboratory measurements and field applications.
It is widely used in the oil industry to evaluate different production processes. Our simulator will be compared against it to show the accuracy.

\subsubsection{Polymer flood in a one-dimensional geometry}

Here we consider a special case of polymer flood in one-dimension geometry. The reservoir contains 15 grid
blocks along the $\xd$-direction where the properties are uniformly distributed at each grid block with porosity
$\porosity = 0.5$ and intrinsic permeability $\perm =100\,mD$. A polymer adsorption curve and water viscosity multiplier are displayed in \Fig{\ref{fig:polymer}}, and the fluid properties are summarized in
\Tab{\ref{table:exam:1}}.

\begin{figure}[!hbt]
  \centering
  \includegraphics[width=\linewidth]{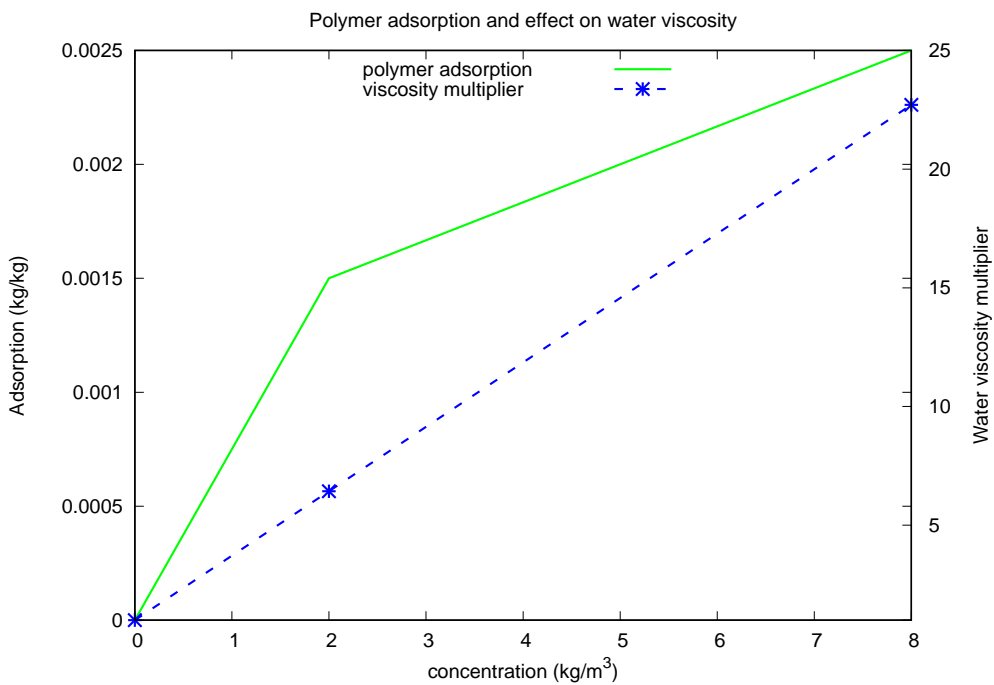}\\
  \caption{Polymer adsorption curve and water viscosity multiplier.}\label{fig:polymer}
\end{figure}

 \begin{table}[!htb]
  \centering
  \caption{Simulation input data of one-dimensional geometry case.}\label{table:exam:1}
  \begin{tabular}{p{.78\linewidth}  c} \\\toprule
  Model parameter & Value\\\midrule
  Spatial grid size ($m$) & 100 \\
  Initial resident water saturation & 0.4 \\
  Initial polymer concentration ($kg/m^3$) & 0.1 \\ \hline
  \multicolumn{2}{c}{Water}\\ \hline
  Mass density ($kg/m^3$) & 1025.18 \\
  Formation volume factor ($sm^3/rm^3$) & 1.0 \\
  Compressibility ($1/Bar$) & 3.03e-6 \\
  Viscosity ($cp$) & 0.5 \\ \hline
  \multicolumn{2}{c}{Oil} \\\hline
  Mass density ($kg/m^3$) & 832.96 \\
  Formation volume factor ($sm^3/rm^3$) & 1.0\\
  Compressibility ($1/Bar$) & 1.0e-5 \\
  Viscosity ($cp$) & 0.5\\  \hline
  \multicolumn{2}{c}{Rock} \\\hline
  IPV & 0.15 \\
  RRF & 2.67 \\
  Maximum polymer adsorption (kg/kg) & 0.0035 \\
  \bottomrule\end{tabular}
\end{table}

We inject water with a fixed constant volume rate at $1000\,m^3/day$ by the injection well located at the first grid block. The production well is perforated through the 15th grid block, where a liquid volume well constraint was operated at $1000\, m^3/day$. The polymer injection concentration varies along the production process. At the beginning, only pure water is injected for $300\, days$, and then polymer is added with concentration at $6\,kg/m^3$ to improve the volumetric sweep efficiency. Finally, we stop injecting polymer after $500\,days$ injection, and simulation is terminated after total $1800\,days$.

In \Fig{\ref{fig:exam:1}}, the oil and water production rates are compared with those from Schulumberger-Eclipse that are represented by the green line, while our numerical solutions are plotted in red color and line markers. After about $500\,days$ of water injection, the water front has broken through, and the oil production rate reduces until the whole reservoir is flooded. A good agreement is found in \Fig{\ref{fig:exam:1}} with some minor differenced due to numerical diffusion and different ways to approximate the water phase viscosity. 

\begin{figure*}[!htb]
  \centering
  \includegraphics[width=\linewidth]{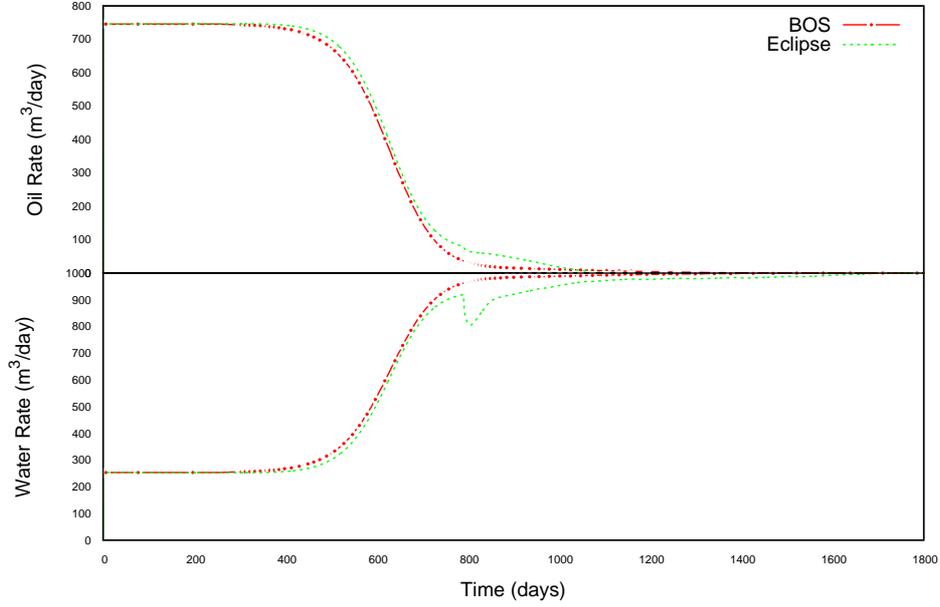}\\
  \caption{Comparison between BOS and Eclipse on 1D example.}\label{fig:exam:1}
\end{figure*}

\subsubsection{Polymer flood in a two-dimensional geometry}
Next, we simulate a polymer flood case on a $10 \times 10$ grid and the formation is initially fully saturated with oil with porosity $\porosity = 0.2$ and intrinsic permeability $\perm =50\,mD$. The fluid properties are summarized in \Tab{\ref{table:exam:2}}.

\begin{table}[!htb]
  \centering
  \caption{Simulation input data on $10 \times 10$ grid.}\label{table:exam:2}
  \begin{tabular}{p{.78\linewidth}  c} \\\toprule
  Model parameter & Value\\\midrule
  Spatial grid size ($ft$) & 75 \\
  Initial polymer concentration ($lbm/bbl$) & 0. \\ \hline
  \multicolumn{2}{c}{Water}\\ \hline
  Mass density ($lbm/cuft$) & 64 \\
  Formation volume factor ($STB/bbl$) & 1.0 \\
  Compressibility ($1/psia$) & 3.03e-6 \\
  Viscosity ($cp$) & 0.5 \\ \hline
  \multicolumn{2}{c}{Oil} \\\hline
  Mass density ($lbm/cuft$) & 52 \\
  Formation volume factor ($STB/bbl$) & 1.0\\
  Compressibility ($1/psia$) & 1.0e-5 \\
  Viscosity ($cp$) & 2\\
  \bottomrule\end{tabular}
\end{table}

There is no polymer in the reservoir initially. An injection well locates at the left corner of the grid and a production well locates at the other end of the diagonal. Water and a polymer solution are injected at a maximum injection rate of $200\, STB/day$ with polymer concentration at $50\, lbm/STB$. At the same time, the injection well is constrained with its bottom hole pressure, with no more than $2\times 10^5\, psia$. There is no constraint on the production rate, but the bottom hole pressure of the production well is fixed at $3999\,psia$. After $200\,days$ production, there is no polymer injected into the formation. The simulation is terminated after total $1700\,days$.

The same case was carried out by Schulumberger-Eclipse, and the oil and water production rates are compared in \Fig{\ref{fig:exam:2}} by the red and green colors that represent our simulator and Eclipse, respectively. The same production pattens are shown with different simulators, and the difference between them is negligible.

\begin{figure*}[!htb]
  \centering
  \includegraphics[width=\linewidth]{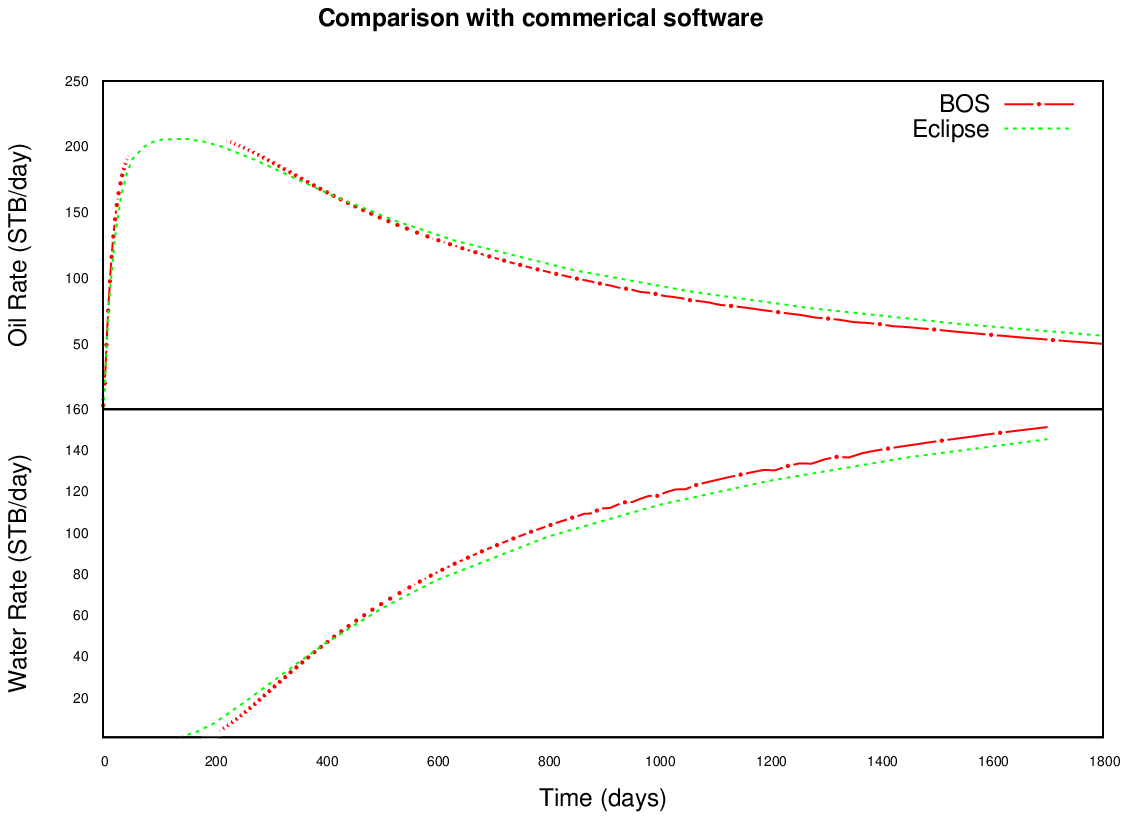}\\
  \caption{Comparison between BOS and Eclipse on a two-dimensional polymer flood.}\label{fig:exam:2}
\end{figure*}

\subsubsection{Polymer flood in a three-dimensional geometry}
In this case, we extend the two-dimensional case into a three-dimensional problem. The same as in the
two-dimensional problem, two vertical wells locate at the two ends of the diagonal of a $xy$ plane and act as
an injection well and a production well. Both wells have a multiple layer perforation that is different from
the two-dimensional case. For multiple perforated layer wells, a reference bottom hole pressure $\pressure_b$
is selected at a datum depth. The fluid density is calculated at that depth and the initial pressure in the
reservoir is determined by marching down the reservoir in small steps by recalculating the density at each
step. It is treated explicitly through the Newton iteration. This algorithm is more complicated when more than
one phase is present. Starting at the datum depth, the hydrostatic pressure for the datum phase can be
calculated by marching up and down the reservoir. Pressures in the other phases can then be determined at the
contact depths, and then the hydrostatic pressures can be determined throughout the reservoir by marching up
and down again. Once the phase pressures are known, the phase saturations can be determined at each depth so
that the hydrostatic pressure variation is balanced by the capillary pressure between the phases.

The agreement of our simulator with commercial software is shown in \Fig{\ref{fig:exam:3}} for the oil and water production rates.

\begin{figure*}[!htb]
  \centering
  \includegraphics[width=\linewidth]{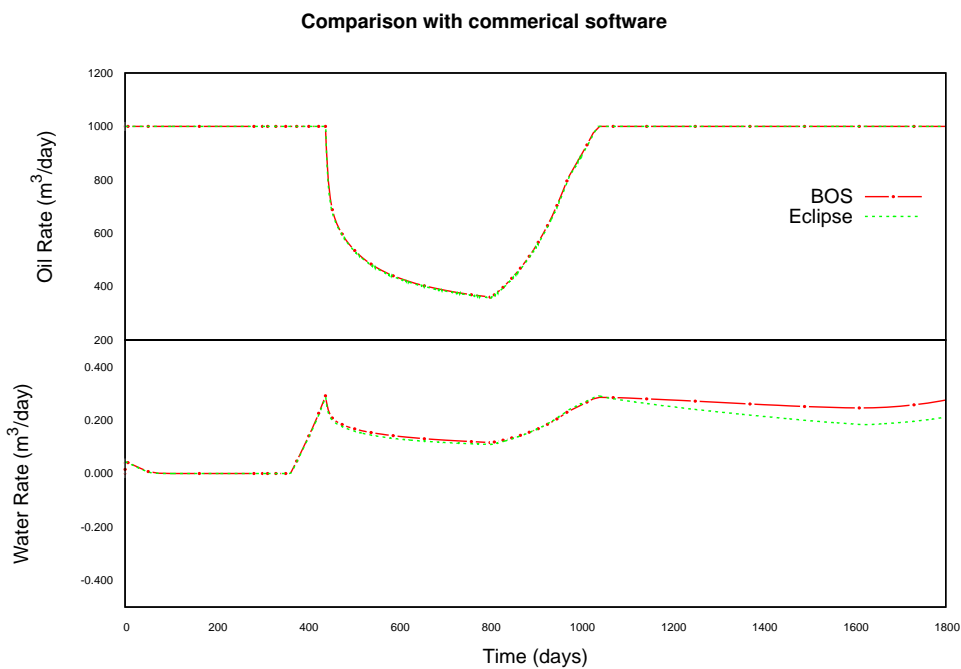}\\
  \caption{Comparison between BOS and Eclipse with multiple layer wells and capillary gravity equilibrium state.}\label{fig:exam:3}
\end{figure*}

\subsection{Performance Test}
In this section, the performance of our simulator is reported in terms of speedup. The speedup, $\speedup$, is defined as the ratio of the elapsed time when executing a simulation by $r$ cores to the execution time when $n$ processors are used:
\begin{equation} \speedup_n = \frac{T_r}{T_n} \end{equation}
Independent timings are performed in different running cases by our simulator. In the ideal scenario, the amount of time reduces linearly with the number of CPUs rising.
From a performance point of view, our simulator is demonstrated to provide significant speedup with respect to multi-core CPUs through our tests.

\subsubsection{Low resolution reservoir model}
The first performance test is carried out on a polymer flooding project with $95\times 192\times 5$ grid blocks in the $\xd$, $\yd$ and $\zd$ directions, respectively, on a IBM System x3750 M4 Intel Xeon processor-based, entry-level 4-socket server. The server consists of 32 3.3GHz SMP CPU cores with 512 GB local RAM and 10GbE networking options in a 2U form factor.

The reservoir properties, such as porosity and permeability, vary along different layers, as \Tab{\ref{table:test:1}} shows. The initial reservoir pressure is $4000\, psi$ at a reference depth of $6150\, ft$. The simulation was based on water flood by a vertical and horizontal well patten for $3980\, days$, followed by polymer flood for almost $360\, days$. Four vertical injection wells are distributed at the corners of the reservoir, while a horizontal producer is drilled at the center of the reservoir.

\begin{table}[!htb]
  \centering
  \caption{Reservoir description}\label{table:test:1}
  \begin{tabular}{c c c c c} \\\toprule
  Layer & Porosity & \multicolumn{3}{c}{Permeability (mD)} \\
  & &  X  & Y  & Z \\\midrule
  1 & 0.17393 & 326.4 & 980.6 & 163.4 \\
  2& 0.1694& 445.3 & 1335.8 & 222.6 \\
  3& 0.25714& 148.9& 446.6& 74.4 \\
  4& 0.17344 & 118.8& 356.4 & 59.4 \\
  5& 0.1187 & 71.2 & 213.6	& 35.6 \\
  \bottomrule\end{tabular}
\end{table}

The results reported in \Fig{\ref{fig:speed:m}} demonstrate the performance of our simulator. Simulations were performed from 32 (the reference) to 256 cores. It is shown in \Fig{\ref{fig:speed:m}} that the simulator approximately has the linear speedup performance.

In the case of water flood, the water phase advances through the heterogeneous formation much faster compared to the polymer flood. This is due to the fact that pure water being less viscous is more mobile. On the other hand, the polymer flood is able to achieve a much more efficient sweep of the reservoir. Although this makes polymer flooding a slower recovery process compared to waterflooding, the net oil recovery can be substantially more in polymer flood. The polymer flood results clearly show that a larger area has been swept and also the higher water saturation levels have penetrated deeper into the domain compared to the water flood.

\begin{figure*}[!htb]
  \centering
  \includegraphics[width=\linewidth]{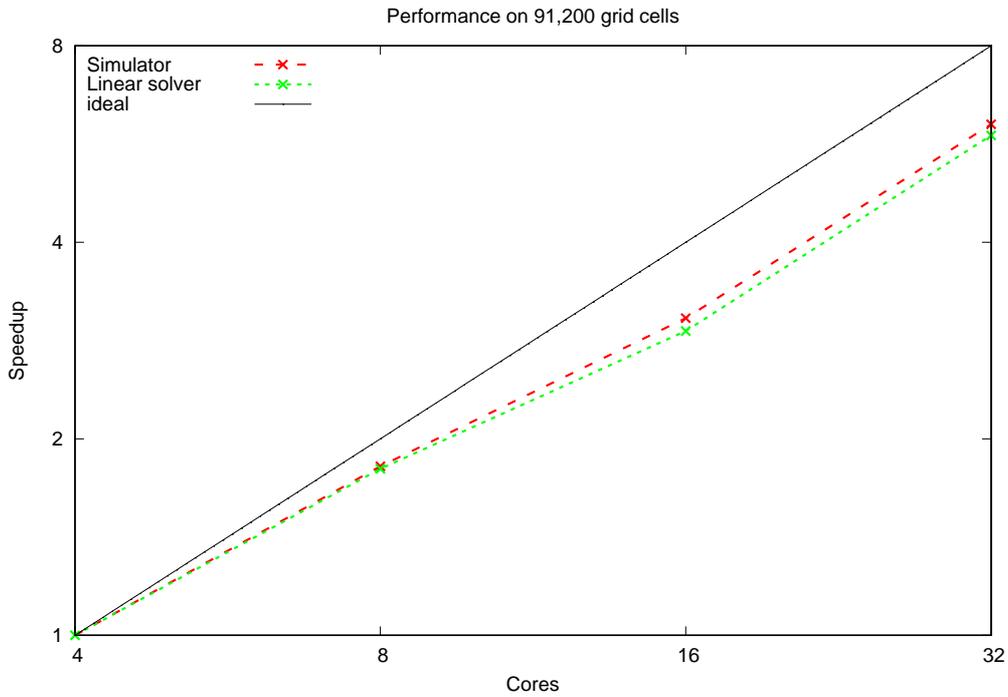}\\
  \caption{Log-log representation of the speedup on Platform and linear solver (reference is 4 cores). }\label{fig:speed:m}
\end{figure*}

\subsubsection{High resolution reservoir model}
The second performance test is carried out by Graham cluster supplied by Compute Canada. The Graham consists of  35,520 cores and 320 GPU devices, spreaded across 1,107 nodes. All nodes except bigmem3000 have Intel E5-2683 V4 CPUs, running at 2.1 GHz. This test was run on the nodes with 128GB memory where a low-latency high-bandwidth Infiniband fabric connects all nodes and scratch storage. Nodes configurable for cloud provisioning also have a 10Gb/s Ethernet network, with 40Gb/s uplink to scratch storage. The same reservoir description as in the last case is used except with the refined grid size of $951\times 1920\times 15$. Simulations were performed from 128 (the reference) to 2048 cores.

This example tests the scalability of the platform as well as the solution of linear systems (including the restarted GMRES iteration solver and the CPR preconditioner). \Fig{\ref{fig:speed:l}} shows that when processors are doubled, the elapsed time costed by the platform and linear solver is cut by half, which means that our simulator has excellent scalability. This example also shows that the simulator can deal with more accurate geological models at multi-million and even multi-billion grid blocks under accepted run time if there are enough computation resources available.

\begin{figure*}[!htb]
  \centering
  \includegraphics[width=\linewidth]{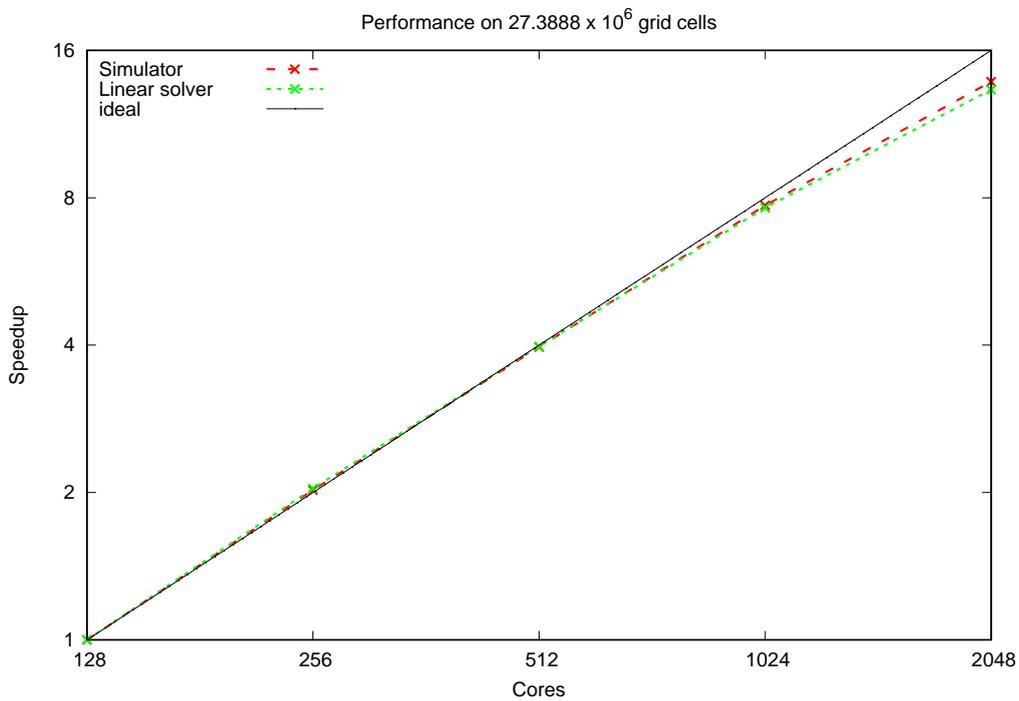}\\
  \caption{Log-log representation of the speedup on Platform and linear solver (reference is 128 cores). }\label{fig:speed:l}
\end{figure*}

\section{Conclusions}
The results of this work demonstrate that coupled fluid flow and polymer adsorption phenomena can be effectively simulated and distributed among multi-core CPUs. High performance computing techniques are applied through our in-house simulator, which has the capability to build and solve large-scale reservoir systems, especially reservoir models with high resolutions and complexity. The rapid advancements in parallel computer hardware and software technologies are adapted to improve the efficiency of polymer flooding processes. A satisfactory parallel efficiency of our simulator is demonstrated for a complex field application with up to 27 million grid blocks by 2048 cores. Almost linear speedup is displayed.

\section{Acknowledgments}
The support of Department of Chemical and Petroleum Engineering, University of Calgary and Reservoir Simulation Group is gratefully acknowledged. The research is partly supported by NSERC/Energi Simulation Chair, Alberta Innovate Chair, IBM Thomas J.Watson Research Center, and the Frank and Sarah Meyer FCMG Collaboration Centre for Visualization and Simulation. The research is also enabled in part by support provided by WestGrid (www.westgrid.ca), SciNet (www.scinethpc.ca) and Compute Canada Calcul Canada (www.computecanada.ca).

\section*{References}

\bibliography{library}

\end{document}